\documentclass
[twocolumn,showpacs,preprintnumbers,floats,amsmath,amssymb]{revtex4}%
\usepackage{graphicx}
\usepackage{dcolumn}
\usepackage{bm}
\usepackage{amsmath}
\usepackage{amsfonts}
\usepackage{color}
\usepackage{amssymb}%
\providecommand{\U}[1]{\protect\rule{.1in}{.1in}}
\begin{document}
\title{Cosmic anisotropic doomsday in Bianchi type I universes}
\author{Mauricio Cataldo}
\altaffiliation{mcataldo@ubiobio.cl}

\author{Antonella Cid}
\altaffiliation{acidm@ubiobio.cl}

\author{Pedro Labra\~na}
\altaffiliation{plabrana@ubiobio.cl}

\affiliation{Departamento de F\'\i sica, Universidad del B\'\i
o--B\'\i o, Avenida Collao 1202, Casilla 5-C, Concepci\'on, Chile,
and\\
Grupo de Cosmolog\'\i a y Gravitaci\'on-UBB}

\author{Patricio Mella}
\altaffiliation{patricio.mella@uach.cl}

\affiliation{Centro de Docencia de Ciencias B\'asicas para
Ingenier\'\i a,
Facultad de Ciencias de la Ingenier\'\i a, Universidad Austral de Chile, Casilla 567, Valdivia, Chile, and \\
Grupo de Cosmolog\'\i a y Gravitaci\'on-UBB}

\date{\today}

\begin{abstract}
\textbf{{Abstract:}} In this paper we study finite time future
singularities in anisotropic Bianchi type I models. It is shown that
there exist future singularities similar to Big Rip ones (which
appear in the framework of phantom {Friedmann-Robertson-Walker}
cosmologies). Specifically, in an ellipsoidal anisotropic scenario
or in a fully anisotropic scenario, the three directional and
average scale factors may diverge at a finite future time, together
with energy densities and anisotropic pressures.  We call these
singularities ``Anisotropic Big Rip Singularities". We show that
{there also exist} Bianchi type I models filled with matter, where
one or two directional scale factors may diverge. Another type of
future anisotropic singularities is shown to be present in vacuum
cosmologies, i.e. Kasner spacetimes. These singularities are induced
by the shear scalar, which {also blows up}  at a finite time. We
call such a singularity ``Vacuum Rip". In this case one directional
scale factor blows up, while the other two and average scale factors
tend to zero.

\vspace{0.5cm}

\end{abstract}

\pacs{98.80.Cq, 04.30.Nk, 98.70.Vc} \maketitle

\section{Introduction}
The astrophysical observations \cite{howell} give evidence that our
Universe is currently in {accelerated} expansion.  In the context of
Einstein General Relativity this acceleration is driven by an
unknown fluid called dark energy \cite{Copeland}, usually described
by a state parameter $w=p/\rho$, with $w<-1/3$. This corresponds to
quintessence matter which violates the strong energy condition, and
the range $w<-1$ to phantom matter, which violates the strong and
dominant energy conditions. In this latter case we could have the
scale factor, $\rho$ and $p$ going to infinity at a finite cosmic
time in the future. This type of singularity is dubbed Big Rip
\cite{Caldwell:2003vq}. This possibility is allowed for isotropic
and homogeneous Friedmann-Robertson-Walker (FRW) models by current
observational data \cite{alam}.

In Big Rip scenarios the curvature invariants $R^{2}$,
$R_{\mu\nu}R^{\mu\nu}$, $R_{\mu\nu\rho\sigma }R^{\mu\nu\rho\sigma}$
diverge in the same way as occur in the Big-Bang and Big-Crunch
singularities \cite{Dabrowskisea,stoica}. However, in the framework
of FRW cosmologies there are different sorts of finite time future
singularities. According to Ref.~\cite{Nojiri,type5} {the future
singularities} can be classified in the following types:

Type $I$ (\textquotedblleft Big Rip\textquotedblright) : For
$t\rightarrow t_{s}$, $a\rightarrow\infty$, $\rho \rightarrow\infty$
and $|p|\rightarrow \infty$.

Type $II$ (\textquotedblleft Sudden\textquotedblright) : For
$t\rightarrow t_{s}$, $a\rightarrow a_{s}$, $\rho
\rightarrow\rho_{s}$ and $|p|\rightarrow\infty$.

Type $III$ (\textquotedblleft Big Freeze\textquotedblright) : For
$t\rightarrow t_{s}$, $a\rightarrow a_{s}$, $\rho \rightarrow\infty$
and $|p|\rightarrow\infty$.

Type $IV$ (\textquotedblleft Generalized sudden\textquotedblright):
For $t\rightarrow t_{s}$, $a\rightarrow a_{s}$, $\rho \rightarrow0$,
$|p|\rightarrow0$ and higher derivatives of $H$ diverge.

Type $V$ (\textquotedblleft w-singularities\textquotedblright): For
$t \rightarrow t_s$, $a \rightarrow a_s$, $\rho \rightarrow 0$, $p
\rightarrow 0$, $w \rightarrow \infty$ and higher derivatives of $H$
are regular.

The quantities  $t_{s}$, $a_{s}$, $\rho_{s}$ and $p_{s}$ are
constants.

The type II singularity has been studied by several authors
\cite{sudden,suddenb,type4b,Dabrowski:2004bz} and includes the
subcases of the Big Brake and Big Boost \cite{bbbb}. In Ref.
\cite{fjl} it was shown that for this type of singularity the
universe can be extended after the singular event. The type III, IV
and V have been studied in Refs. \cite{type3},
\cite{Dabrowskisea,Nojiri} and \cite{type5}, respectively.

There are other types of future singularities that can appear at a
finite time, even when the strong energy condition is satisfied
\cite{fjl,barrow,suddenb,bbbb}. Other interesting types of future
singularities, but appearing at an infinite time, are \emph{Little
Rip} \cite{lr},\emph{ Pseudo-Rip} \cite{sr} and \emph{Little Sibling
Rip of Big Rip} \cite{srbr}. It is noteworthy to mention that an
attempt to unify future singular behaviors was made in Ref.
\cite{grcr}, where the authors introduce the Grand Rip and Grand
Bang/Crunch singularities.

It is interesting to note that phantom fields are not the only way
to generate scenarios with Big Rip. Such future singularities may be
induced, for instance, by fluids with an inhomogeneous equation of
state \cite{ieos} or interacting coupled fluids \cite{cfr}. From the
viewpoint of viscous cosmological models, in Ref. \cite{cat} it was
shown that the bulk viscosity induces a Big Rip singularity, and in
Ref.~\cite{bre} it was studied a Little Rip as a purely viscous
effect. Inhomogeneous and spherically symmetric gravitational
fields, describing evolving wormholes also may exhibit a Big Rip
singularity during its evolution~\cite{wcat}. Notice that in
Ref.~\cite{Dabrowski:2004bz} an anisotropic and inhomogeneous
cosmology of Stephani type was found to possess finite-time sudden
singularities, and in Ref.~\cite{type4b} specific examples of
anisotropic sudden singularities in Bianchi type VII$_0$ universes
{were} constructed.

In this paper we extend the study of future singularities by
considering anisotropic and homogeneous spacetimes more general than
flat FRW ones. Specifically, we analyze Bianchi type I cosmologies,
allowing us to show that the anisotropy of spacetime, by means of
the shear scalar, may induce future singularities at a finite time,
similar to Big Rip ones appearing in the framework of phantom FRW
cosmologies. In order to make analytical progress on this topic we
shall use some known exact Bianchi type I solutions of the Einstein
equations, allowing us to handle exact expressions for directional
scale factors $a_i(t)$, shear scalar $\sigma$, energy density $\rho$
and anisotropic pressures $p_i$.

Our motivation is based on the fact that several studies on the
plausibility of anisotropy in the accelerated expanding universe
have been performed in the framework of anisotropic dark energy
cosmological models. In Ref.~\cite{Campanelli} authors found that,
in the framework of Bianchi I cosmological models, anisotropy is
permitted both in the geometry of the universe and in the dark
energy equation of state. Additionally, it is worth to mention that
an anisotropic dark energy model can potentially solve the CMB
low-quadrupole problem~\cite{Rodriguez}.

The paper is organized as follows. In Sec. II we write the Einstein
equations for Bianchi type I spacetimes. In Sec. III we discuss the
Kasner vacuum solution and the future singularities which may appear
during its evolution. In Sec. IV we find future singularities in
anisotropic Bianchi type I models filled with a stiff fluid. In Sec.
V we obtain exact solutions for ellipsoidal universes characterized
by a shear scalar proportional to the expansion scalar, and filled
with isotropic and anisotropic matter sources. We show that these
spacetimes may exhibit anisotropic rip singularities. In Sec. VI we
discuss future singularities in fully anisotropic Bianchi I
cosmologies filled with an anisotropic, matter source. In Sec. VII
we discuss our results.

\section{Bianchi type I spacetimes and Einstein Field Equations}
In this paper we consider models belonging to spatially homogeneous
and anisotropic Bianchi type I spacetimes described by the metric
\begin{eqnarray}\label{BI metric}
ds^2=dt^2-a_1^2(t) dx^2-a_2^2(t) dy^2-a_3^2(t) dz^2,
\end{eqnarray}
where $a_i(t)$ are the directional scale factors along the $x, y, z$
axes, respectively.

This type of cosmologies is {particularly interesting} because it is
the simplest generalization of the homogeneous and isotropic flat
FRW models.

The Einstein field equations for this metric may be written in the
following form~\cite{Chimento}:
\begin{eqnarray} \label{E1}
3H^2=\kappa \rho+\frac{\sigma^2}{2}, \\ \label{E2}
-2 \dot H=\kappa (\rho + p) +\sigma^2, \\ \label{E3}
\dot \rho+ 3 H (\rho+p) = \vec \sigma \cdot \vec \Sigma, \\
\label{E4}
\dot{ \vec{\sigma}}+3H \vec \sigma=\vec \Sigma,
\end{eqnarray}
where $\kappa=8 \pi G$, {we will consider $\kappa=1$ from here on}.
The average expansion rate $H$, the average pressure $p$, the shear
vector $\vec \sigma$, and the transverse pressure vector $\vec
\Sigma$ are respectively defined as
\begin{eqnarray}\label{B1}
H=\frac{1}{3} \left(H_1+H_2+H_3 \right), \\ \label{B2} p=\frac{1}{3}
\left(p_1+p_2+p_3 \right), \\ \label{B3} \sigma_i=H_i-H, \\
\label{B4} \Sigma_i=p_i-p,
\end{eqnarray}
where $i=1,2,3$. The average expansion rate may be written as
$H=\dot{\bar{a}}/\bar{a}$, where the average scale factor $\bar a$
is defined by
\begin{equation}
\bar{a}=(a_1 a_2 a_3)^{1/3},
\end{equation}
and the directional expansion rates are given by
\begin{equation}
H_i=\frac{\dot{a}_i}{a_i}.
\end{equation}
From Eqs.~(\ref{B1})-(\ref{B4}) we see that the quantities $\vec
\sigma$ and $\vec \Sigma$ satisfy the constraints
\begin{eqnarray}
\sigma_1+\sigma_2+\sigma_3=0, \\
\Sigma_1+\Sigma_2+\Sigma_3=0,
\end{eqnarray}
respectively.

Additionally we give the definition of some useful anisotropic
quantities. The shear tensor $\sigma_{a b}$ is defined by
\begin{eqnarray*}
\sigma_{ab}=h_a^c u_{(c;d)} h_b^d -\frac{1}{3} \theta h_{ab},
\end{eqnarray*}
where $\theta=u^c_{; c}$ is the expansion scalar, $h_{ab}= g_{ab} -
u_a u_b$ the projection tensor (for the signature $(+, -, -, -)$),
and $u_a$ the four-velocity. From this expression we obtain the
shear scalar given by $\sigma^2= \sigma_{a b}\sigma^{a b}$.

For the considered Bianchi type I metric~(\ref{BI metric}) we have
that the expansion scalar, non-zero shear tensor components and the
shear scalar are given by
\begin{eqnarray*}
\theta=H_1+H_2+H_3, \\
\sigma_{1}^1=  -\frac{2}{3} H_1+\frac{1}{3} \left(H_2+H_3
\right), \\
\sigma_{2}^2=-\frac{2}{3}
H_2+\frac{1}{3} \left(H_1+H_3 \right), \\
\sigma_{3}^3= -\frac{2}{3} H_3+\frac{1}{3} \left(H_1+H_2 \right),
\\ \sigma^2=\frac{2}{3} \left(H_1^2+H_2^2+H_3^2-H_1 H_2-H_1 H_3 -H_2
H_3 \right),
\end{eqnarray*}
respectively.

\section{Finite-time future anisotropic singularities in vacuum Kasner spacetimes}
In this section we study future singularities of anisotropic
character {by considering} Bianchi type I spacetimes without matter,
i.e. vacuum solutions for the metric~(\ref{BI metric}) (or Kasner
spacetimes).

By putting $\rho=0$ and $p_i=0$ into Eqs.~(\ref{E1})-(\ref{E4}) we
obtain the following four independent differential equations:
\begin{eqnarray}\label{K1}
3H^2=\frac{\sigma^2}{2}, \\ \label{K2} \dot \sigma_i+3H \sigma_i=0.
\end{eqnarray}
From Eq.~(\ref{K2}) we obtain
\begin{equation}\label{KK1}
\sigma_i=\frac{\sigma_{i0}}{\bar{a}^3},
\end{equation}
then
$\sigma^2=(\sigma^2_{10}+\sigma^2_{20}+\sigma^2_{30})/\bar{a}^6$ and
Eq.~(\ref{K1}) gives
\begin{equation}\label{KK2}
\bar{a}(t)=\left(C \pm \frac{{1}}{2} \sqrt{6} \, \sigma_0  \, t
\right)^{1/3}.
\end{equation}
Here $\sigma_{i0}$ and $C$ are {integration constants}, and
\begin{equation}\label{15}
\sigma_0 \equiv \sqrt{2(\sigma^2_{10}+\sigma^2_{20}+\sigma_{10}
\sigma_{20})},
\end{equation}
where we have used the relation
\begin{equation}
\sigma_{30}=-\sigma_{10}-\sigma_{20}.
\end{equation}

From Eqs.~(\ref{B3}), (\ref{K1}) and~(\ref{KK1}) we may write
relation
\begin{eqnarray}\label{relacionentreHs}
\frac{\dot{a}_i}{a_i}=\left(1 \pm
\frac{\sqrt{6}\sigma_{i0}}{\sigma_0} \right)
\frac{\dot{\bar{a}}}{\bar{a}},
\end{eqnarray}
which implies that the directional expansion rates $H_i$ are
proportional to the average expansion rate $H$. By using
Eq.~(\ref{KK2}) we obtain for directional scale factors
\begin{equation}\label{losai}
a_i=a_{i0}^{\pm} \, \left(C \pm\frac{{1}}{2} \sqrt{6} \, \sigma_0 \,
t \right)^{\frac{1}{3} \left(1 \pm \frac{\sqrt{6} \,
\sigma_{i0}}{\sigma_0}\right)},
\end{equation}
where $a_{i0}^{\pm}$ are integration constants for branches $+$ and
$-$ respectively, and $i=1,2,3$.

In order to proceed with the analysis, we shall use the initial
condition $H_1(t=0)=H_0>0$ for the directional scale factor $a_1$.
This implies that at $t=0$ we are imposing an expanding scale factor
$a_1$. Then, from Eq.~(\ref{losai}) we obtain that $C= \frac{6
\sigma_{10} \pm \sqrt{6}\sigma_0}{6H_0}$, and the scale factor along
$x$-direction takes the form
\begin{eqnarray}
a_1=a_{10}^\pm \left( 1+ \frac{3 H_0 \, t}{1 \pm  \frac{\sqrt{6}
\sigma_{_{10}}}{\sigma_{_0}} }  \right)^{\frac{1}{3} \left( 1 \pm
\frac{\sqrt{6} \sigma_{_{10}}}{\sigma_{_0}} \right)}.
\end{eqnarray}

Then, the metric~(\ref{BI metric}) is given by
\begin{eqnarray}
ds^2=dt^2-\left( 1+ \frac{3 H_0 \, t}{1 \pm  \frac{\sqrt{6}
\sigma_{_{10}}}{\sigma_{_0}} } \right)^{\frac{2}{3} \left(1
\pm\frac{\sqrt{6} \, \sigma_{10}}{\sigma_0}\right)} dx^2- \nonumber
\\ \left( 1+ \frac{3 H_0 \, t}{1 \pm  \frac{\sqrt{6}
\sigma_{_{10}}}{\sigma_{_0}} } \right)^{\frac{2}{3} \left(1 \pm
\frac{\sqrt{6} \, \sigma_{20}}{\sigma_0}\right)} dy^2 - \nonumber \\
\left( 1+ \frac{3 H_0 \, t}{1 \pm  \frac{\sqrt{6}
\sigma_{_{10}}}{\sigma_{_0}} } \right)^{\frac{2}{3} \left(1 \mp
\frac{\sqrt{6} \, (\sigma_{10}+\sigma_{20})}{\sigma_0}\right)} dz^2,
\label{BI metric Kasner}
\end{eqnarray}
where the constants $a_{i0}$ have been absorbed by rescaling the
spatial coordinates.

Since we are interested in solutions with finite time future
singularities one should require that $1 \pm \frac{\sqrt{6} \,
\sigma_{10}}{\sigma_0}<0$. This implies that for the positive branch
of the considered solution the condition
\begin{eqnarray}\label{condi1}
\sigma_{10} < -\frac{\sigma_0}{\sqrt{6}} <0
\end{eqnarray}
must be fulfilled by coefficients $\sigma_{10}$ and $\sigma_{20}$,
while for the negative branch the condition
\begin{eqnarray}\label{condi2}
\sigma_{10} > \frac{\sigma_0}{\sqrt{6}}>0
\end{eqnarray}
must be required. Therefore, in metric~(\ref{BI metric Kasner}) the
scale factor along $x$-direction exhibits a future singularity at
finite value of the cosmic time $t_{vr}=-\frac{1}{3H_0} \left (1 \pm
\frac{\sqrt{6} \sigma_{10}}{\sigma_0}  \right)>0$. Notice that
Eq.~(\ref{relacionentreHs}) implies that in this scenario we have
that $H_1>0$ and $H<0$. If we demand that the scale factor along
$y$-direction also becomes singular at the finite value $t_{vr}$ we
must require additionally that $1 \pm \frac{\sqrt{6} \,
\sigma_{20}}{\sigma_0}<0$, which implies that $\sigma_{20} <
-\frac{\sigma_0}{\sqrt{6}} <0$ for the positive branch, and
$\sigma_{20} > \frac{\sigma_0}{\sqrt{6}}>0$ for the negative branch.
However, it can be shown that simultaneously it is not possible to
satisfy the conditions  $\sigma_{10} < -\frac{\sigma_0}{\sqrt{6}}$
and $\sigma_{20} < -\frac{\sigma_0}{\sqrt{6}}$ (or $\sigma_{10} >
\frac{\sigma_0}{\sqrt{6}}$ and $\sigma_{20} >
\frac{\sigma_0}{\sqrt{6}}$).

\begin{figure}[ht!]
\includegraphics[scale=0.55]{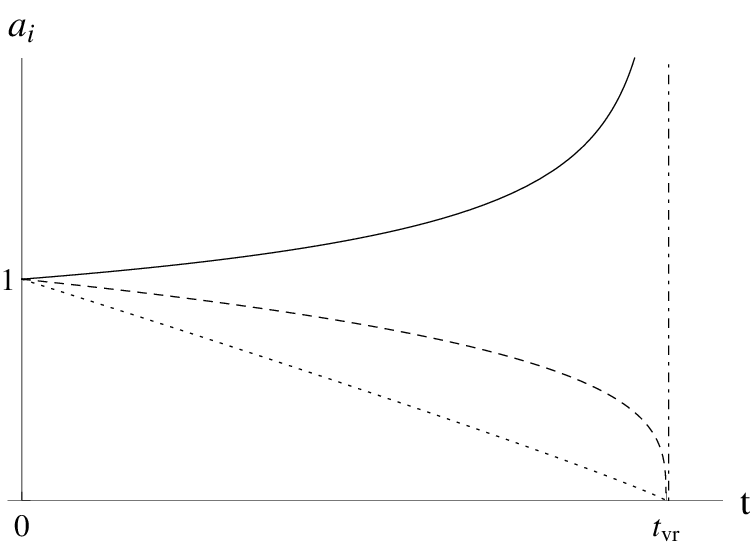}
\caption{The figure shows the qualitative behavior of the scale
factors $a_i$ for fully anisotropic ($a_1 \neq a_2 \neq a_3$) vacuum
Bianchi type I solutions with $\sigma_{10}$ and $\sigma_{20}$
satisfying Eqs.~(\ref{condi1}) or~(\ref{condi2}). In the figure are
shown two of the scale factors ($a_2$ and $a_3$) which at different
rates of contraction tend to zero at the vacuum rip time $t_{vr}$
(dotted and dashed lines), while the third one, $a_1$, diverges at
this time (solid line). In this case the future singularity is of
anisotropic Cigar Rip type (see TABLE~\ref{Table15}).} \label{Fig15}
\end{figure}

Therefore, only the scale factor $a_1$ becomes singular at finite
value of the cosmic time $t_{vr}=-\frac{1}{3H_0} \left (1 \pm
\frac{\sqrt{6} \sigma_{10}}{\sigma_0}  \right)>0$, while the other
two scale factors $a_2$ and $a_3$ become zero at this time (see
Fig.~\ref{Fig15}). It is interesting to note that for all
directional scale factors~(\ref{losai}), the corresponding expansion
rates $H_i$ diverge at $t_{vr}$. From Eq.~(\ref{relacionentreHs}) we
conclude that the same is valid for the average expansion rate $H$.

It is worth to mention that if we consider $C=0$ in Eq.(\ref{KK2}),
then we have $C=0$ in Eq.(\ref{losai}) and the directional scale
factor $a_1$ diverges at $t_{vr}=0$ instead of
$t_{vr}=-\frac{1}{3H_0} \left (1 \pm \frac{\sqrt{6}
\sigma_{10}}{\sigma_0}  \right)$. The condition for the occurrence
of this singularity is the same as before
$1\pm\sqrt{6}\sigma_{10}/\sigma_0<0$. In this sense, the occurrence
of the singularity is independent of the value of the constant $C$:
if $C=0$ we must consider $t<0$ for the consistency of
Eq.(\ref{KK2}), i.e. $C\pm\sqrt{6}\sigma_0t/2>0$, if $C\neq0$ the
consistency of Eq.(\ref{KK2}) $C\pm\sqrt{6}\sigma_0t/2>0$ allows us
to consider $t_{vr}<0$ or $t_{vr}>0$.

Notice that by defining
\begin{eqnarray}\label{kk15}
q_1=\frac{1}{3} \left(1 \pm \frac{\sqrt{6} \,
\sigma_{10}}{\sqrt{2(\sigma^2_{10}+\sigma^2_{20}+\sigma_{10}\sigma_{20})}}\right), \nonumber \\
q_2=\frac{1}{3} \left(1 \pm \frac{\sqrt{6} \,
\sigma_{20}}{\sqrt{2(\sigma^2_{10}+\sigma^2_{20}+\sigma_{10}\sigma_{20})}}\right),  \\
q_3=\frac{1}{3} \left(1 \mp \frac{\sqrt{6} \, (\sigma_{10}+
\sigma_{20})}{\sqrt{2(\sigma^2_{10}+\sigma^2_{20}+\sigma_{10}\sigma_{20})}}\right),
\nonumber
\end{eqnarray}
for the powers of the scale factors in Eq.~(\ref{BI metric Kasner}),
we obtain that
\begin{eqnarray}\label{KC1}
q_1+q_2+q_3=1, \\
q^2_1+q^2_2+q^2_3=1,\label{KC2}
\end{eqnarray}
where $q_1$, $q_2$ and $q_3$ are the Kasner parameters. These
constraints {correspond to} the conditions for {the} well known
vacuum Kasner solution.

\begin{figure}[ht!]
\includegraphics[scale=0.55]{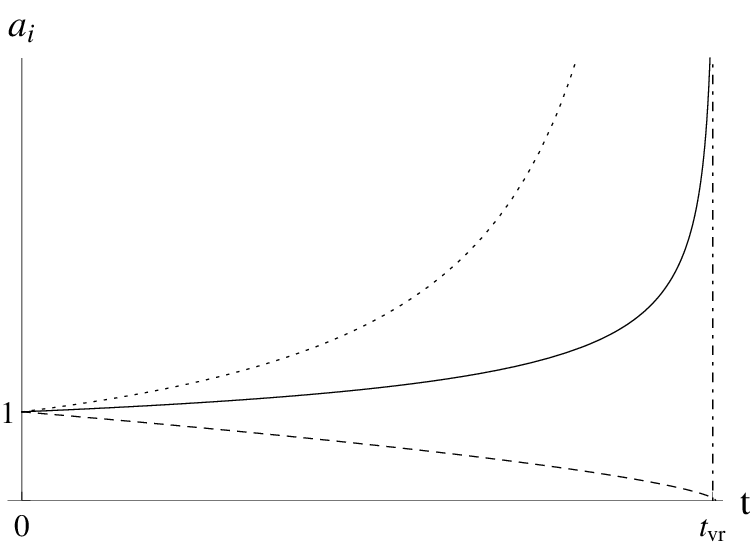}
\caption{The figure shows the qualitative behavior of the scale
factors $a_i$ for an ellipsoidal vacuum solution given by
$\sigma_{10}=\sigma_{20}=-1$, or equivalently by $q_1=q_2=2/3$ and
$q_3=-1/3$. In this case, two of the scale factors at the same
contraction rate tend to zero at $t_{vr}$ (dashed line), and the
other scale factor blows up at this time (solid line). Here we have
included the qualitative behavior of any of the directional
expansion rates $H_i$ (dotted line), which diverges at $t_{vr}$. In
this case the future singularity is of symmetric Cigar Rip type (see
TABLE~\ref{Table15}).} \label{Fig15A}
\end{figure}

From the Kasner conditions~(\ref{KC1}) and~(\ref{KC2}) we note that
if we arrange the Kasner parameters in increasing order
$q_1<q_2<q_3$, then they change in the ranges~\cite{Belinski}
\begin{eqnarray}
-\frac{1}{3} \leq q_1  \leq 0, \\
0 \leq q_2  \leq \frac{2}{3}, \\
\frac{2}{3} \leq q_3  \leq 1.
\end{eqnarray}
From these relations we conclude again that if it is present a
future singularity only one of the scale factors may blow up, while
the other two tend to zero at a finite value of the cosmic time. For
the particular case of an ellipsoidal vacuum cosmology the following
parameter values must be required: $q_1=q_2=0,q_3=1$ or $q_1=-1/3,
\, q_2=q_3=2/3$. Therefore, only the latter set of parameter values
allow us to have a future singularity for an ellipsoidal vacuum
universe (see Fig.~\ref{Fig15A}).

In conclusion, due to the anisotropic character of Bianchi type I
metrics, in the Kasner vacuum solution all three scale factors do
not exhibit simultaneously a future singularity: just one of the
scale factors may exhibit such a singularity at $t_{vr}$, while the
other two do not. In this case the average scale factor~(\ref{KK2})
does not exhibit a singular behavior, becoming zero at
$t_{vr}=-\frac{1}{3H_0} \left (1 \pm \frac{\sqrt{6}
\sigma_{10}}{\sigma_0}  \right)$. We note that this scenario
necessarily corresponds to an average contracting universe. However,
all directional expansion rates $H_i$ as well as the average
expansion rate $H$ diverge at $t_{vr}$, and due to the vacuum
character of the Kasner solutions, the scalar curvature is always
zero.

Because of the absence of matter content, the discussed anisotropic
future singularities are not similar to any of the finite-time
singularities listed in the introduction section. We shall call such
a singularity a Vacuum Rip.

It should be emphasized that these vacuum rips are not produced by
fluids violating the dominant energy conditions
(DEC)~\cite{Hawking}, i.e. $\rho \geq 0$ and $-p \leq \rho \leq p$,
as stated for FRW cosmologies filled with a phantom fluid. The
Kasner vacuum solution satisfies DEC, and by writing the Kasner
metric in the form where the shear is explicitly included, we have
shown that future singularities may be induced by the anisotropy of
the spacetime, by making a suitable choice of the model parameters
$\sigma_{10}$ and $\sigma_{20}$.

\section{Finite-time future anisotropic singularities with a stiff fluid}

To elucidate the role of the shear, in the occurrence of future
singularities, in the presence of matter fields we will consider the
``toy model" of fully anisotropic Bianchi type I spacetimes~(\ref{BI
metric}), filled with a stiff fluid, for which the condition for the
powers of scale factors~(\ref{KC1}) is still valid. This ``toy
model" is interesting because it allows us to consider finite time
future singularities in a cosmology, fulfilling energy conditions
($p=\rho$, $\rho \geq 0$). The other aspect to be considered is that
this cosmological model allows us to handle exact analytical
expressions for studying relevant quantities.

We use the field equations in the form given by
Eqs.~(\ref{E1})-(\ref{E4}). Since the pressure is isotropic, in this
case we have that $\vec{\Sigma}=0$, and from Eq.~(\ref{E3}) we have
for the equation of state $p=\rho$ that the energy density is given
by $\rho=\rho_0/\bar{a}^6$. On the other hand, from Eq.~(\ref{E4})
we obtain the solution~(\ref{KK1}), and then Eq.~(\ref{E1}) implies
that the average scale factor is given by
\begin{equation}\label{B15}
\bar{a}(t)=\left( C\pm\sqrt{3\rho_0+\frac{3}{2} \sigma_0^2} \, t
\right)^{1/3},
\end{equation}
where $C$ is an integration constant and $\sigma_0$ is given by
Eq.~(\ref{15}). Note that by making $\rho_0=0$ we obtain the average
scale factor~(\ref{KK2}) discussed in the previous section.


From Eq.~(\ref{B3}) we may write for directional expansion rates
$H_i$ the following equation:
\begin{eqnarray}
{\frac {\dot{a}_{{i}} }{a_{{i}}  }} \mp \frac{1}{3}\,{\frac {\sqrt
{12\,\rho_{{0}}+6\,{\sigma_{{0}}}^{2}}} {2\,C \pm \sqrt
{12\,\rho_{{0}}+6\,{\sigma_{{0}}}^{2}} \, t}}= \\ \nonumber
 {\frac {\sigma_{ {i0}}}{C \pm \sqrt
{3\,\rho_{{0}}+\frac{3}{2}\,{\sigma_{{0}}}^{2}} \,t}},
\end{eqnarray}
which implies that the directional scale factors are given by
\begin{eqnarray}\label{SFani15}
a_i=a_{i0}^\pm \left(C \pm \frac{\sqrt{12 \rho_0+6\sigma_0^2}}{2} \,
t \right)^{\frac{1}{3} \pm \frac{2\sigma_{i0}}{\sqrt{12 \rho_{0}+6
\sigma_0^2}}},
\end{eqnarray}
where $a_{i0}^\pm$ are {integration constants}.

By using the initial condition $H_1(t=0)=H_0>0$ for the directional
scale factor $a_1$ we obtain from Eq.~(\ref{SFani15}) that
$C=\frac{6 \sigma_{10}\pm\sqrt{12 \rho_0+6 \sigma_0^2}}{6H_0}$, and
then the scale factor along $x$-direction takes the form
\begin{eqnarray}
a_1=a_{10}^\pm \left(1+\frac{3\sqrt{12 \rho_0+6\sigma_0^2}}{\sqrt{12
\rho_0 + 6\sigma_0^2} \pm6\sigma_{10}} H_0 \, t\right)^{\frac{1}{3}
\pm \frac{2\sigma_{10}}{\sqrt{12 \rho_0+6
\sigma_0^2}}}. \nonumber \\
\end{eqnarray}

Hence, the resulting metric may be written as
\begin{eqnarray}\label{metrica final stiff f}
ds^2=dt^2-\left(1+ \frac{H_0}{\gamma} \, t\right)^{\frac{2}{3} \pm
\frac{4\sigma_{10}}{\sqrt{12 \rho_0+6 \sigma_0^2}}} dx^2- \nonumber
\\ \left(1+ \frac{H_0}{\gamma} \,
t\right)^{\frac{2}{3} \pm \frac{4\sigma_{20}}{\sqrt{12 \rho_0+6
\sigma_0^2}}}dy^2 - \nonumber \\ \left(1+ \frac{H_0}{\gamma} \,
t\right)^{\frac{2}{3} \mp \frac{4(\sigma_{10}+\sigma_{20})}{\sqrt{12
\rho_0+6 \sigma_0^2}}} dz^2,
\end{eqnarray}
and the energy density and the pressure are given by
\begin{eqnarray}
\rho=p=\frac{36 H_0^2 \rho_0}{\left(\sqrt{12 \rho_0 +6 \sigma_0^2}
\pm 6\sigma_{10}\right)^2 \left(1+ \frac{H_0}{\gamma} \, t
\right)^2},
\end{eqnarray}
where
\begin{eqnarray}
\gamma=\frac{1}{3} \pm \frac{2\sigma_{10}}{\sqrt{12 \rho_0+6
\sigma_0^2}},
\end{eqnarray}
i.e. the power of the scale factor along $x$-direction.

In order to induce a future singularity, and considering that
$H_0>0$, we must require that $\gamma<0$. This implies that
\begin{equation}
\sigma_{10} < - \frac{1}{6} \, \sqrt{12 \rho_0+6\sigma_0^2}<0,
\end{equation}
for the positive branch, and
\begin{equation}
\sigma_{10} > \frac{1}{6} \, \sqrt{12 \rho_0+6\sigma_0^2}>0,
\end{equation}
for the negative branch.

By taking into account Eq.~(\ref{15}) we conclude that
\begin{equation}\label{15as}
\sigma_{10} <  \frac{1}{4} \left(\sigma_{20}- \sqrt{9 \,
\sigma_{20}^2+8 \, \rho_0}\right) <0,
\end{equation}
for the positive branch, and
\begin{equation}\label{15bs}
\sigma_{10} >  \frac{1}{4} \left(\sigma_{20}+ \sqrt{9 \,
\sigma_{20}^2+8 \, \rho_0}\right) >0,
\end{equation}
for the negative branch.

Notice that by defining the powers of the scale factors in the
metric~(\ref{metrica final stiff f}) as
\begin{eqnarray}\label{kk15a}
q_1=\frac{1}{3} \pm \frac{2\sigma_{10}}{\sqrt{12 \rho_0+6
\sigma_0^2}}, \nonumber \\
q_2=\frac{1}{3} \pm \frac{2\sigma_{20}}{\sqrt{12 \rho_0+6 \sigma_0^2}},  \\
q_3=\frac{1}{3} \mp \frac{2(\sigma_{10}+\sigma_{20})}{\sqrt{12
\rho_0+6 \sigma_0^2}}, \nonumber
\end{eqnarray}
the parameters $q_i$ satisfy the condition~(\ref{KC1}),
independently of the values of $\sigma_{10}$, $\sigma_{20}$ and
$\rho_0$. Then, the average scale factor takes the form
\begin{eqnarray}\label{averageSFSM}
\bar a=\left(1+\frac{3\sqrt{12 \rho_0+6\sigma_0^2}}{\sqrt{12
\rho_0+6\sigma_0^2} \pm 6\sigma_{10}} H_0 \, t\right)^{\frac{1}{3}}.
\end{eqnarray}
However, now we have that
\begin{eqnarray}\label{KKCC2}
q_1^2+q_2^2+q_3^2 \neq1.
\end{eqnarray}
By putting $\rho_0=0$ the parameters $q_i$ in Eqs.~(\ref{kk15a})
become the Kasner parameters of relations~(\ref{kk15}), implying
that the condition~(\ref{KC2}) is fulfilled for vanishing matter.

Therefore, in the case of Bianchi type I cosmologies filled with
stiff matter the singularity appears at
$t_{rs}=-\frac{\gamma}{H_0}$. As in the vacuum case, from
Eqs.~(\ref{kk15a}) we note that we can have only one of the
directional scale factors blowing up together with the energy
density and pressure. It becomes clear from
expressions~(\ref{metrica final stiff f})-(\ref{15as}) that this rip
singularity is induced by the anisotropy of {the} spacetime.

\begin{figure}[ht!]
\includegraphics[scale=0.55]{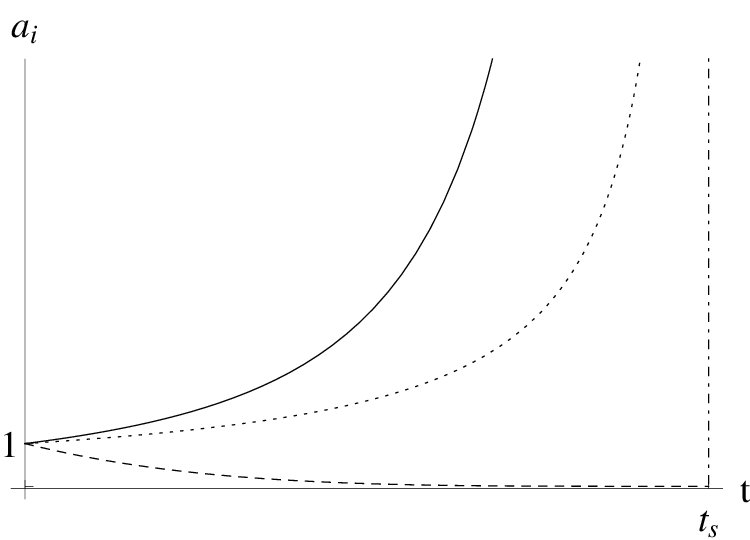}
\caption{The figure shows the qualitative behavior of the scale
factors $a_i$ for fully anisotropic Bianchi type I solutions with
equation of state $p=\rho$ and satisfying the condition~(\ref{KC1})
with $q_1<0$ and $q_2<0$. Here two of the scale factors evolve at
different rates of expansion and diverge at $t_s$ (solid and dotted
lines), while the third scale factor tends to zero at this time
(dashed line). Note that {in this case} $\rho<0$ as shown in
FIG.~\ref{Fig15Ascalarriccidensidadapromedio}. In this case the
future singularity is of pancake rip type (see
TABLE~\ref{Table15}).} \label{Fig15Ascalefactors}
\end{figure}

\begin{figure}[ht!]
\includegraphics[scale=0.55]{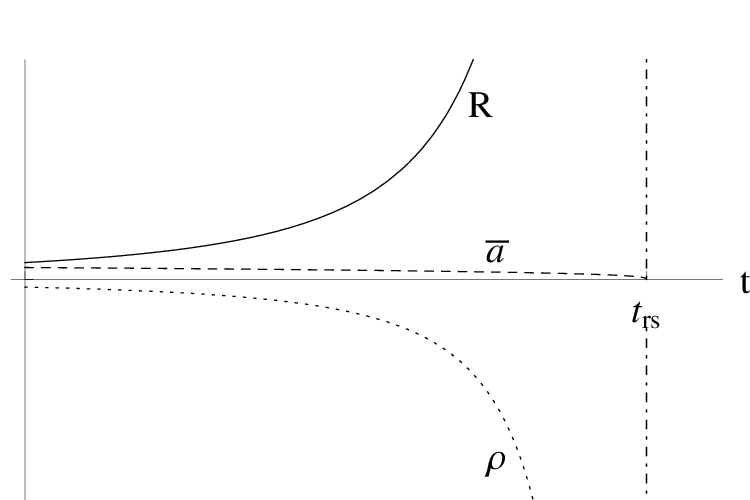}
\caption{The figure shows the qualitative behavior of the energy
density (doted line), scalar curvature (solid line) and the average
scale factor (dashed line) {for fully anisotropic Bianchi type I
solutions with stiff matter}. In this case, the energy density is
negative, and together with the Ricci scalar blow up at $t_s$, while
the average scalar factor tends to zero.}
\label{Fig15Ascalarriccidensidadapromedio}
\end{figure}

Note that, if we require that two of powers $q_i$ are negative, as
it is allowed by the condition~(\ref{KC1}), then  $\rho_0<0$,
implying that the energy density becomes negative, violating the
weak energy condition (see Figs.~\ref{Fig15Ascalefactors}
and~\ref{Fig15Ascalarriccidensidadapromedio}).

The average scale factor~(\ref{averageSFSM}) does not exhibit a
singularity at $t_{rs}$, where it vanishes. This anisotropic rip
singularity appears at $t_{rs}$ only for a 
contracting average scale factor.

\section{Future anisotropic rip singularities in ellipsoidal universes}
In this section we consider the evolution of Bianchi type I
cosmologies with a matter content characterized by isotropic and
anisotropic pressure.

Specifically, we consider particular cases of Bianchi type I models
described by the condition $a_1(t)=a_2(t)$. Thus the line
element~(\ref{BI metric}) takes the form
\begin{eqnarray}\label{Emetric}
ds^2=dt^2-a_1^2(t) (dx^2+dy^2)-a_3^2(t) dz^2,
\end{eqnarray}
which possesses spatial sections with planar symmetry and an axis of
symmetry directed along the $z$-axis. The functions of the cosmic
time $a_1(t)$ and $a_3(t)$ are the directional scale factors along
x, y and z directions respectively. The metric~(\ref{Emetric})
describes a space that has an ellipsoidal rate of expansion at any
moment of the cosmological time, dubbed also Locally Rotationally
Symmetric Bianchi I.

In this case the Einstein field equations~(\ref{E1})-(\ref{E4}) may
be written in the form
\begin{eqnarray} \rho=\frac{\dot{a}_1^2}{a_1^2}+2
\frac{\dot{a}_1 \, \dot{a}_3}{a_1 \,
a_3}, \label{00}\\
p_1=-\left( \frac{\ddot{a}_1}{a_1} + \frac{\dot{a}_1 \,
\dot{a}_3}{a_1
\, a_3} +\frac{\ddot{a}_3}{a_3} \right), \label{11}\\
p_3= - \left( 2 \frac{\ddot{a}_1}{a_1}+\frac{\dot{a}_1^2}{a_1^2}
\right), \label{22}
\end{eqnarray}
where $p_1=p_2$ and $p_3$ are the transversal and longitudinal
pressures respectively. For the metric~(\ref{Emetric}) the average
scale factor is given by $\bar{a}(t)=(a_1^2(t) \, a_3(t))^{1/3}$,
and the average {expansion rate} takes the form
\begin{eqnarray}\label{mean H}
H=\frac{\dot{\bar{a}}}{\bar{a}}=\frac{1}{3} \left( 2
\frac{\dot{a}_1}{a_1}+\frac{\dot{a}_3}{a_3} \right).
\end{eqnarray}

In order to handle exact solutions to the metric~(\ref{Emetric}), we
further make the assumption that the scale factors $a_1$ and $a_3$
are constrained to be given by
\begin{eqnarray}\label{a(t)b(t)}
a_1(t)= a^{\alpha}_3(t),
\end{eqnarray}
where $\alpha$ is a constant. Thus the metric~(\ref{Emetric}) takes
the following form:
\begin{eqnarray}\label{holographic metric}
ds^2=dt^2-a_3^{2\alpha}(t) (dx^2+dy^2)-a_3^2(t) dz^2.
\end{eqnarray}
This metric becomes isotropic for $\alpha=1$. It is interesting to
note that the metric~(\ref{holographic metric}) is characterized by
the condition that expansion scalar $\theta=(2+\alpha) H$ is
proportional to the shear scalar $\sigma^2$.

The measure of anisotropy $\sigma/\theta$ is constant in a number of
Bianchi-type spacetimes representing perfect fluid cosmologies with
barotropic equations of state (see~\cite{Roy} and references
therein). Given that these models may allow nearly isotropic
scenarios, they can be used for studying the effects of anisotropy
in our universe by confronting them with observational data. In our
case, this condition will allow us to work with analytical
ellipsoidal cosmologies exhibiting anisotropic rip singularities.


\subsection{Anisotropic rip singularities with isotropic pressure: $p=\rho$}
Let us suppose that $p_1=p_2=p_3=p$. Thus from
Eqs.~(\ref{00})-(\ref{22}) and~(\ref{a(t)b(t)}), the relevant metric
function $a_3(t)$ is given by
\begin{eqnarray}\label{SFb}
a_3(t)= c_1 \left( 1 +c_2  \, t\right)^\frac{1}{(2\alpha+1)},
\end{eqnarray}
where $c_1$ and $c_2$ are integration constants.

We shall rewrite this scale factor by using for the directional
Hubble parameter $H_3$ the condition
\begin{equation}\label{initial condition}
H_3(t=0)=H_{0}.
\end{equation}
Thus, the scale factor~(\ref{SFb}) takes the form
\begin{eqnarray}\label{SFbH0}
a_3(t)=c_1 \left(1 +(2 \alpha+1)H_0 \, t
\right)^\frac{1}{(2\alpha+1)},
\end{eqnarray}
and the metric~(\ref{holographic metric}) is given by
\begin{eqnarray}\label{holographic isotropic metric}
ds^2=dt^2-\left(1 +(2 \alpha+1)H_0 \, t \right)^{\frac{2 \alpha}{2
\alpha+1}} (dx^2+dy^2) \nonumber \\ -\left(1 +(2 \alpha+1)H_0 \, t
\right)^\frac{2}{2 \alpha+1} dz^2,
\end{eqnarray}
where the constant $c_1$ has been absorbed by rescaling the spatial
coordinates.

In this case the energy density and pressure result to be:
\begin{eqnarray}\label{stiff pressure}
\rho=p=\frac{\alpha(\alpha+2)H_0^2}{  \left(1 +(2 \alpha+1)H_0 \, t
\right)^2},
\end{eqnarray}
which means that the isotropic requirement for the pressure implies
that the matter filling the universe is characterized by a stiff
equation of state.

From the expression~(\ref{SFbH0}) we see that a future rip
singularity appears for $\alpha<-1/2$ when $H_0>0$, or
$-1/2<\alpha<0$ for $H_0<0$. On the other hand, in order to have a
positive energy density we must also require that $\alpha <-2$ or
$\alpha>0$, which excludes the case $H_0<0$. For $H_0>0$, the scale
factor~(\ref{SFbH0}), energy density and pressure blow up at the
finite value of the cosmic time $t_{rs}=-\frac{1}{(2\alpha+1)H_0}$,
while the scale factor $a_1=a_3^\alpha$ becomes zero at this time.
In this case the average scale factor is given by
\begin{eqnarray}\label{averageSFbH0}
\bar{a}(t)= \left(1 +(2 \alpha+1)H_0 \, t \right)^\frac{1}{3},
\end{eqnarray}
and does not exhibit a singularity at $t_{rs}$.

In conclusion, for the metric~(\ref{holographic metric}) the
requirement of isotropic pressure implies that the matter content
behaves as a stiff fluid. The evolution of this cosmology exhibits a
future singularity for $\alpha<-2$ ($H_0>0$) at
$t_{rs}=-\frac{1}{(2\alpha+1)H_0}$. Due to the functions $a_3(t)$,
$\rho(t)$ and $p(t)$ blow up at this time, this singularity is
similar to the FRW Big Rip one but of anisotropic character, since
at $t_{rs}$ the scale factor along $x$ and $y$ directions becomes
zero, as well as the average scale factor $\bar{a}$. From
Eq.~(\ref{averageSFbH0}) we note that this scenario corresponds to a
contracting universe.

As in Sec. III, the anisotropic future singularities are not
produced by fluids violating the DEC, since in this case $\rho=p$.
This type of singularities is induced by the anisotropy of the
spacetime, since if shear vanishes, i.e. $\sigma=0$, then the
solution becomes the standard isotropic FRW cosmology filled with a
stiff fluid, which does not exhibit any future singularity at a
finite value of the cosmic time, and only presents the initial
singularity or Big Bang.

\begin{figure}
\includegraphics[scale=0.55]{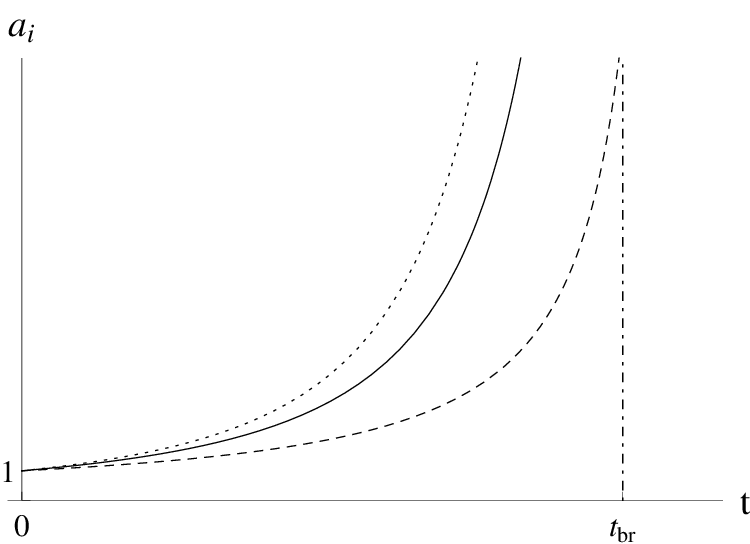}
\caption{The figure shows the qualitative behavior of the scale
factors $a_i$ (dashed and dotted lines) and the average scale factor
(solid line) for ellipsoidal universes with anisotropic pressure.
All them blow up at $t_{br}$.} \label{Fig15AAA}
\end{figure}

\begin{figure}
\includegraphics[scale=0.55]{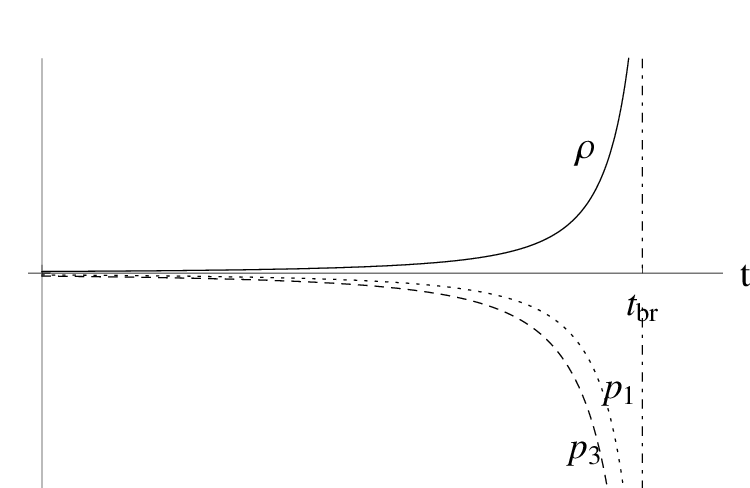}
\caption{The figure shows the qualitative behavior of the energy
density (solid line) and anisotropic pressures $p_1=p_2$ and $p_3$
(dotted and dashed lines respectively). In this case, all quantities
diverge at $t_{br}$.} \label{Fig15AAAA}
\end{figure}

\subsection{Big Rip singularities with anisotropic pressure}
Let us suppose that the {transversal and longitudinal} pressures are
given by
\begin{eqnarray}
p_1=\omega_1 \rho, \\
p_3=\omega_3 \rho,
\end{eqnarray}
respectively, where $\omega_1$ and $\omega_3$ are constant state
parameters. Thus, from Eqs.~(\ref{00}),~(\ref{11})
and~(\ref{a(t)b(t)}) we obtain that
\begin{eqnarray}\label{SFani}
a_3(t)=c_1(1+c_2 \, t)^{\frac{\alpha+1}{\alpha^2 \omega_1
+\alpha^2+2\alpha \omega_1 +\alpha+1}},
\end{eqnarray}
where $c_1$ and $c_2$ are integration constants. By using the
initial condition~(\ref{initial condition}) the scale
factor~(\ref{SFani}) takes the form
\begin{eqnarray}\label{SFani final}
a_3(t)=c_1 \left(1+ \frac{H_0}{\gamma} \, t\right)^{\gamma},
\end{eqnarray}
where
\begin{eqnarray}\label{gamma}
\gamma=\frac{\alpha+1}{\alpha^2 \omega_1 +\alpha^2+2\alpha \omega_1
+\alpha+1}.
\end{eqnarray}
{In this case }the metric takes the form:
\begin{eqnarray}\label{metrica final}
ds^2=dt^2-\left(1+ \frac{H_0}{\gamma} \, t\right)^{\alpha \gamma} (dx^2+dy^2) - \nonumber \\
\left(1+ \frac{H_0}{\gamma} \, t\right)^{\gamma} dz^2,
\end{eqnarray}
where the constant $c_1$ has been absorbed by rescaling the spatial
coordinates and we have considered $H_3(t=0)=H_0$, and the energy
density and the pressure $p_3$ take the form
\begin{eqnarray}
\rho=\frac{\alpha (\alpha+2)H_0^2}{\left(1+ \frac{H_0}{\gamma} \,
t\right)^2},
\end{eqnarray}
\begin{eqnarray}\label{pp22}
p_3=\frac{1+2\alpha \omega_1-\alpha}{1+\alpha} \, \rho,
\end{eqnarray}
respectively. Eq.~(\ref{pp22}) implies that the state parameter
$\omega_3$ is given by
\begin{eqnarray}\label{omega22}
\omega_3 =\frac{1+2\alpha \omega_1-\alpha}{1+\alpha}.
\end{eqnarray}

It becomes clear that, {in order to have a positive energy density
we must require $\alpha<-2$ or $\alpha>0$ and} for $\gamma<0$ the
scale factor~(\ref{SFani final}) exhibits a future singularity at
$t_{rs}=-\frac{\gamma}{H_0}$. At this value of the cosmic time the
energy density and pressures also blow up. In this case for
$\alpha<-2$ we can have one of the scale factors blowing up at
$t_{rs}$, while for $\alpha >0$ all three scale factor may blow up
at $t_{rs}$. It is interesting to note that the average scale
factor, given by
\begin{eqnarray}\label{a medio final}
\bar{a}(t)=\left(1+ \frac{H_0}{\gamma} \, t\right)^{\frac{\gamma(2
\alpha+1)}{3}},
\end{eqnarray}
also may exhibit a singular behavior at $t_{rs}$ for $\gamma<0$ and
$\alpha>0$. These inequalities imply, with the help of
Eq.~(\ref{gamma}), that $\omega_1<-\frac{1+\alpha+\alpha^2}{2 \alpha
+\alpha^2}$, hence we have that the state parameter $\omega_1$ can
not be greater than $- \sqrt{3}/2$ for any $\alpha>0$. For
$\alpha<-2$ the power of Eq.~(\ref{a medio final}) is always
positive and the average scale factor tends to zero at the time
$t_{rs}$, while the scale factors $a_1$ and $a_2$ go to zero and
$a_3$ blows up at this time. This singularity corresponds to an
axisymmetric pancake rip defined in TABLE~\ref{Table15}. Note that
for $\alpha>0$ we have $H>0$, and for $\alpha<-2$ we have $H<0$.

In conclusion, in the case of ellipsoidal universes we may have one,
or two, or all three scale factors blowing up at
$t_{rs}=-\frac{\gamma}{H_0}$. For the latter case, it is crucial to
require $\alpha >0$ and $\alpha^2 \omega_1 +\alpha^2+2\alpha
\omega_1 +\alpha+1<0$. This condition will be realized by requiring
\begin{eqnarray}
\alpha > 1 \,\,\,\,\, \textrm{for} \,\,\,\,\, \omega_1=-1, \\
\alpha_-<\alpha<\alpha_+ \,\,\,\,\, \textrm{for} \,\,\,\,\,
-1<\omega_1<-\frac{\sqrt{3}}{2}, \nonumber  \\
\alpha >\alpha_- \,\,\,\,\, \textrm{for} \,\,\,\,\, \omega_1 < -1.
\end{eqnarray}
where
\begin{equation}
\alpha_{\pm}=\frac{1}{2}\frac{-(1+2\omega_1) \pm
\sqrt{4\omega_1^2-3}}{1+\omega_1}.
\end{equation}
Therefore, we can have future singularities of cigar and pancake rip
types (see TABLE~\ref{Table15}). The cigar singularities may be of
anisotropic ($a_1 \rightarrow \infty, a_2 \rightarrow 0, a_3
\rightarrow 0$) or symmetric types ($a_1 \rightarrow \infty, a_2=a_3
\rightarrow 0$), while the pancake singularities may be anisotropic
and infinite ($a_1 \rightarrow \infty, a_2 \rightarrow \infty$ and
$a_3 \rightarrow 0$) or of axisymmetric (infinite) type ($a_1= a_2
\rightarrow \infty, a_3 \rightarrow 0$).

In Figs.~\ref{Fig15AAA} and~\ref{Fig15AAAA} we show the qualitative
behaviors of the three scale factors, average scale factor, energy
density and pressures for ellipsoidal universes~(\ref{metrica
final}). In {this} example the future singularity is of Big Rip
type, and the universe rips itself apart in all directions at a
finite time, with diverging energy density and pressures.

\begin{widetext}
\begin{center}
\begin{table}[h!]
\begin{center}
\begin{tabular}{|l|l|l|l|c|c|}\hline
\multicolumn{2}{|c|}{Initial
Singularities}&\multicolumn{4}{|c|}{Anisotropic Rip
Singularities}\\\hline
    \multicolumn{0}{|c|}{ Type}    & \multicolumn{0}{|c|}{Directional scale factors}       &  \multicolumn{0}{|c|}{Type}        &   \multicolumn{0}{|c|}{Directional scale factors}   &\multicolumn{0}{|c|}{$\sigma$} & \multicolumn{0}{|c|}{$\rho,\vert p_i\vert$} \\\hline
Axisymmetric & $a_1=a_2\rightarrow0,\ a_3\rightarrow0$& & & &\\
Point-like    & & & & &\\\hline
Anisotropic  & $a_1\rightarrow0,\ a_2\rightarrow0,\ a_3\rightarrow0$& & &  &\\
Point-like    & & & & & \\ \hline
Symmetric    & $a_1\rightarrow\textrm{const.},\ a_2=a_3\rightarrow0$& & & &\\
Barrel       & & & & &\\\hline
Anisotropic  & $a_1\rightarrow\textrm{const.},\ a_2\rightarrow0,\ a_3\rightarrow0$& & & & \\
Barrel       & & & & &\\\hline Symmetric    &
$a_1\rightarrow\infty,\ a_2=a_3\rightarrow0$ &  Symmetric  &
$a_1\rightarrow\infty,\
a_2=a_3\rightarrow0$ & $\infty$& $0$ or $\infty$\\
Cigar        & &Cigar  Rip   &   & &\\\hline
Anisotropic  & $a_1\rightarrow\infty,\ a_2\rightarrow0,\ a_3\rightarrow0$ &Anisotropic  &$a_1\rightarrow\infty,\ a_2\rightarrow0,\ a_3\rightarrow0$ & $\infty$& $0$ or $\infty$\\
Cigar        & & Cigar Rip  & & &\\\hline
Axisymmetric & $a_1\rightarrow0,\ a_2=a_3\rightarrow  $ const.  & Axisymmetric &$a_1\rightarrow0,\ a_2=a_3\rightarrow\infty$ & $\infty$& $\infty$\\
Pancake&  &Pancake Rip & & &
\\\hline
Anisotropic & $a_1\rightarrow0,\ a_2\rightarrow\infty,\ a_3\rightarrow\infty$ & Anisotropic &$a_1\rightarrow0,\ a_2\rightarrow\infty,\ a_3\rightarrow\infty$ &$\infty$&$\infty$\\
Pancake  & & Pancake Rip &  & & \\ \hline
                      & & Axisymmetric& $a_1\rightarrow\infty,\ a_2=a_3\rightarrow\infty$&$\infty$&$\infty$\\
                      & & Big Rip& & & \\\hline
                      & & Anisotropic& $a_1\rightarrow\infty,\ a_2\rightarrow\infty,\ a_3\rightarrow\infty$&$\infty$&$\infty$\\
                      & & Big Rip& & & \\\hline
\end{tabular}

\caption{\label{Table15} In this table all possible initial
singularities for Bianchi type I spacetimes are listed
(see~\cite{Bicak}). For comparison, we also list anisotropic rip
type singularities reached {at} a finite time {and described} by
solutions discussed in this paper. We include the directional scale
factors $a_i$, shear scalar $\sigma$, energy density $\rho$ and
pressures $p_i$. Notice that for vacuum rip singularities only
symmetric and anisotropic cigar rip types are possible, and only in
these cases $\rho$ and $\vert p \vert \rightarrow 0$.}

\end{center}
\end{table}
\end{center}

\end{widetext}

\section{Big rip in fully anisotropic Bianchi type I cosmologies}
It is possible to construct a Bianchi type I generalization of the
ellipsoidal cosmology, exhibiting a future singularity, with three
different scale factors and barotropic anisotropic pressures. For
instance, let us choose the scale factors in the form
\begin{eqnarray}
a_i=\left(1+\frac{H_0 t}{\gamma}\right)^{s_i},
\end{eqnarray}
where $s_1=\alpha \gamma$, $s_2=\beta \gamma$ and $s_3=\gamma$,
$\alpha$ and $\beta$ are constants. In this case, from Einstein
equations the energy density and pressures are given by
\begin{eqnarray}
\rho= \frac{(\alpha+\beta+\alpha \beta)H_0^2}{\left(1+\frac{H_0
t}{\gamma} \right)^2},
\end{eqnarray}
and $p_i= \omega_i \rho$ ($i=1,2,3$), where
\begin{eqnarray} \label{omega15}
\omega_1&=&\frac{1+\beta-\gamma(1+\beta^2+\beta)}{\gamma(\alpha+\beta+\alpha
\beta)}, \\
\omega_2&=&\frac{1+\alpha-\gamma(1+\alpha+\alpha^2)}{\gamma(\alpha+\beta+\alpha
\beta)} ,\\
\omega_3&=&\frac{\alpha+\beta-\gamma
(\alpha^2+\beta^2+\alpha\beta)}{\gamma(\alpha+\beta+\alpha \beta)}.
\label{omega15a}
\end{eqnarray}
In this case the average scale factor is given by
\begin{eqnarray}\label{apromediobr}
\bar{a}=\left(1+\frac{H_0 t}{\gamma}\right)^{\frac{(\alpha+\beta+1)
\gamma}{3}}.
\end{eqnarray}
We notice that for $\alpha=\beta$ we {recover} the ellipsoidal
cosmology of Subsection VB.

Now we are interested in studying scenarios with $\gamma<0$ and a
positive energy density. This means that, we must require
\begin{eqnarray}\label{inequacionbrdensidad}
\alpha+\beta+\alpha \beta>0.
\end{eqnarray}
Simultaneously we require that the power of the average scale factor
in Eq.~(\ref{apromediobr}) be negative, i.e.
\begin{eqnarray}\label{inequacionbr}
(\alpha+\beta+1) \gamma<0.
\end{eqnarray}
From Eqs.~(\ref{inequacionbrdensidad}) and~(\ref{apromediobr}) we
obtain
\begin{eqnarray}\label{inecuacionesbr}
\alpha>-1, \,\,\, \beta>-\frac{\alpha}{1+\alpha}.
\end{eqnarray}

It becomes clear that for $\gamma<0$ at
$t_{br}=-\frac{\gamma}{H_0}>0$ a future anisotropic singularity is
present. The character of the singularity depends on the values of
the constants $\alpha$ and $\beta$. From
inequations~(\ref{inecuacionesbr}) 
we obtain two possibilities: three divergent directional scale
factors for $\alpha>0$, $\beta>0$ and two divergent directional
scale factors for $\alpha>0$, $\beta<0$ or $-1<\alpha<0$, $\beta>0$.

For positive $\alpha$ and $\beta$ we have $a_1 \rightarrow \infty$,
$a_2 \rightarrow \infty$, $a_3 \rightarrow \infty$, $\rho
\rightarrow \infty$ and $\vert p_i \vert \rightarrow \infty$. This
type of singularity corresponds to an anisotropic Big Rip in
TABLE~\ref{Table15}. For $\alpha>0$, $\beta<0$ or $-1<\alpha<0$,
$\beta>0$ the singularity corresponds to an anisotropic Pancake Rip
in TABLE~\ref{Table15}, where we also have $\rho \rightarrow \infty$
and $\vert p_i \vert \rightarrow \infty$.

On the other hand, from Eqs.~(\ref{omega15})-(\ref{omega15a}) we
conclude that for positive $\alpha$ and $\beta$  the state
parameters $\omega_1$, $\omega_2$, $\omega_3$ are always negative.
In particular, it is possible to have a phantom anisotropic
cosmology since all $\omega_i<-1$, and then $p_i/\rho<-1$.
Nevertheless, we can not  simultaneously have three state parameters
in the quintessence range $-1<\omega_i<-1/3$, but we can have one
phantom state parameter and the other two ones in the quintessence
range.

In the case of the anisotropic Pancake Rip, the three state
parameters can be simultaneously positive or in the phantom range
but all the state parameters can not be simultaneously in the
quintessence range. In this scenario we have two branches: (a)
$\alpha>0,\  \beta<0$ and $\gamma<0$, where we can have one phantom
state parameter and two state parameters in the quintessence range;
and (b) $-1<\alpha<0,\  \beta>0$ and $\gamma<0$, where we can have
only one state parameter in the quintesence range and the other
parameters could be both positives or one positive and one in the
phantom range.

\section{Final Remarks}
We have probed that Bianchi type I cosmologies may evolve to
finite-time singularities, which have anisotropic character, since
one, or two or even all three scale factors blow up at a finite
future time. These anisotropic singularities have certain
similarities with that of Big Rip type appearing in the framework of
homogeneous and isotropic phantom FRW cosmologies, in which scale
factor, energy density and pressure become infinite at finite future
time.

By considering specific examples illustrating types of future
anisotropic singularities occurring in the framework of Einstein
Bianchi I cosmologies, we study the behavior of directional scale
factors $a_i(t)$, shear scalar $\sigma$, energy density $\rho$ and
anisotropic pressures $p_i$. We show that future singularities may
be induced by the anisotropy of the spacetime.

In the case of vacuum solutions, i.e. Kasner cosmologies, only one
of the scale factors may exhibit such a singular behavior at a
finite cosmic time $t_{vr}$. The other two ones tend to zero at this
time. In other words, in the direction of the scale factor which
blows up at $t_{vr}$ it follows a super-accelerated expansion until
it hits the vacuum rip singularity. We call this type of
singularities vacuum rip. In {Figs}.~\ref{Fig15} and~\ref{Fig15A} we
show the behavior of the scale factors for fully anisotropic and
ellipsoidal vacuum universes. For vacuum rips we have only
singularities of anisotropic cigar rip type ($a_1 \rightarrow
\infty, a_2 \rightarrow 0, a_3 \rightarrow 0$) or symmetric cigar
rip type ($a_1 \rightarrow \infty, a_2=a_3 \rightarrow 0$), see
TABLE~\ref{Table15}.

On the other hand, for non-vacuum Bianchi type I spacetimes the
scale factors also may evolve to a finite-time singularity,
following a super-accelerated expansion until the universe reaches
an anisotropic rip singularity at which the directional and average
Hubble rates, together with the shear scalar, energy density and
pressure of the universe diverge. We have shown that the anisotropy
of spacetime, by means of the shear scalar, may induce such future
singularities.

Notice that this result depends on the initial condition we choose
for the selected directional scale factor $a_i$ at a time $t_0$ (we
have chosen $i=1$ and $t_0=0$). By using the condition
$H_1(t_0)=H_0$ the integration constant of the directional scale
factor $a_1$ is fixed, therefore expanding and contracting scale
factors are included in these anisotropic scenarios. In this work we
chose an increasing directional scale factor in order to get an
infinite directional scale factor at a finite time $t_s>t_0$.

To elucidate the role of the shear, in the occurrence of future
singularities, in the presence of matter fields we have considered
the ``toy model" of fully anisotropic Bianchi type I
spacetimes~(\ref{BI metric}), filled with a stiff fluid, for which
the condition for the powers of scale factors~(\ref{KC1}) is still
valid. This ``toy model" is interesting because it allows us to
consider cosmology, fulfilling energy conditions ($p=\rho$, $\rho
\geq 0$). The other aspect to be considered is that this
cosmological model allows us to handle exact analytical expressions
for relevant physical quantities. The fulfillment of the weak energy
condition implies that only one of the three scale factors may
exhibit such a finite-time singularity. Thus, as for the vacuum rip,
we can have only singularities of anisotropic cigar rip type ($a_1
\rightarrow \infty, a_2 \rightarrow 0, a_3 \rightarrow 0$) or
symmetric cigar rip type ($a_1 \rightarrow \infty, a_2=a_3
\rightarrow 0$). Accordingly, for $\rho=p>0$, the behavior of scale
factors is similar as shown in Figs.~\ref{Fig15} and~\ref{Fig15A}.
In Figs.~\ref{Fig15Ascalefactors}
and~\ref{Fig15Ascalarriccidensidadapromedio} we show the qualitative
behavior of the solution with two negative powers $q_1$ and $q_2$,
which allows to have negative energy density and isotropic pressure.

In the case of ellipsoidal universes filled with matter
characterized by an anisotropic pressure, we may have one, or two,
or all three scale factors blowing up at
$t_{rs}=-\frac{\gamma}{H_0}$. Therefore, we can have future
singularities of cigar and pancake rip types (see
TABLE~\ref{Table15}). The cigar singularities may be of anisotropic
($a_1 \rightarrow \infty, a_2 \rightarrow 0, a_3 \rightarrow 0$) or
symmetric types ($a_1 \rightarrow \infty, a_2=a_3 \rightarrow 0$),
while the pancake singularities may be anisotropic and infinite
($a_1 \rightarrow \infty, a_2 \rightarrow \infty$ and $a_3
\rightarrow 0$) or of axisymmetric (infinite) type ($a_1= a_2
\rightarrow \infty, a_3 \rightarrow 0$). In Figs.~\ref{Fig15AAA}
and~\ref{Fig15AAAA} we show the qualitative behavior of the three
scale factors, average scale factor, energy density and pressures
for ellipsoidal universes~(\ref{metrica final}). In {this} example
the future singularity is of Big Rip type since the universe rips
itself apart in all directions at a finite time, with diverging
energy density and pressures.

It is worth to mention that for a directional scale factor $a_i(t)$,
exhibiting a singularity at the finite time $t_s$, we have that
$a_i(t) \rightarrow 0$ for $t \rightarrow - \infty$, while if a
directional scale factor vanishes at this finite time $t_s$, then
$a_i(t) \rightarrow \infty$ for $t \rightarrow - \infty$. All
suitable initial singularities appearing for $t \rightarrow -\infty$
are included in the TABLE~\ref{Table15}. From this, it becomes clear
that the chosen time $t=0$ has no physical meaning as initial time
and only denotes a particular moment of the cosmological time during
the evolution of {the} considered cosmological models, for which we
have that $-\infty<t \leq t_s$. In other words, the energy density
$\rho$, anisotropic pressures $p_i$ and specific scale factors $a_i$
blow up {at} a time $t_s$ from this particular time $t=0$.

%

We have constructed also a Bianchi type I generalization of the
ellipsoidal cosmology discussed above, exhibiting a future
singularity, with three different scale factors and barotropic
anisotropic pressures: $a_1 \rightarrow \infty$, $a_2 \rightarrow
\infty$, $a_3 \rightarrow \infty$, $\rho \rightarrow \infty$ and
$p_i \rightarrow \infty$ at $t_{br}$. This type of singularity
corresponds to an anisotropic Big Rip in TABLE~\ref{Table15}. Note
that in this last case the average scale factor {and the average
expansion rate} also blow up at $t_{br}$.

Finally, we have shown that it is possible to classify finite time
future singularities present in Bianchi type I models in terms of
the evolution of directional scale factors. This is done in an
analogous way as it is done for Bianchi type I models in the case of
initial singularities \cite{Bicak}. The studied singularities are
present when one, two or the three directional scale factors blow up
at finite time. We note that an anisotropic Big Rip singularity
(where all the directional scale factors, the average scale factor,
the average expansion rate, the density and the pressures blow up at
finite time) is presented in the case of ellipsoidal universes with
anisotropic pressures and in the case of a fully anisotropic
universe with barotropic anisotropic pressures. Both scenarios
correspond to an expanding average scale factor, which could
represent our universe when a small degree of anisotropy  is
considered. We have investigated this last possibility and the
results will be published elsewhere.

\section{Acknowledgements}
This work was supported by CONICYT through Grants FONDECYT N$^0$
1140238 (MC), 11110507 (AC) and 3130444 (PM). It also was supported
by Direcci\'on de Investigaci\'on de la Universidad del B\'\i
o-B\'\i o through the grants GI 121407/VBC, GI 150407/VC (MC, AC,
PL, PM), 140807 4/R (MC), 151307 3/R (AC) and 141407 3/R (PL).


\begin{thebibliography}{999}
\bibitem{howell}  Y. Wang,``Dark energy'', Wiley-vch 2010, ISBN: 978-3-527-40941-9; ``Dark energy:
Observational and theoretical approaches'', edited by Pilar
Ruiz-Lapuente, Cambridge University Press 2010, ISBN: 9781107647022.
\bibitem{Copeland} E.~J.~Copeland, M.~Sami and S.~Tsujikawa, Int.\ J.\ Mod.\ Phys.\ D {\bf 15}, 1753 (2006)  [hep-th/0603057].


\bibitem{Caldwell:2003vq}
  R.~R.~Caldwell, M.~Kamionkowski and N.~N.~Weinberg, Phys.\ Rev.\ Lett.\  {\bf 91},
  071301 (2003)[astro-ph/0302506], Phys. Lett. B \textbf{545}, 23 (2002) [astroph/9908168];
  A. A. Starobinsky, Grav. Cosmol. \textbf{6}, 157 (2000) [astroph/9912054];
  S. M. Carroll, M. Hoffman and M. Trodden, Phys. Rev.D \textbf{68}, 023509 (2003) [astro-ph/0301273];
  L. P. Chimento and R. Lazkoz, Phys. Rev. Lett. \textbf{91}, 211301 (2003) [gr-qc/0307111];
  M. P. D¸abrowski, T. Stachowiak and M. Szyd lowski, Phys. Rev. D \textbf{68}, 103519 (2003) [hep-th/0307128];
  P. F. Gonzalez-Diaz, Phys. Lett. B \textbf{586}, 1 (2004) [astroph/0312579], Phys. Rev. D \textbf{69}, 063522 (2004)
  [hep-th/0401082]; L.~Fernandez-Jambrina and R.~Lazkoz, Phys.\ Rev.\ D {\bf 74}, 064030 (2006)[gr-qc/0607073].



\bibitem{alam}   U. Alam, V. Sahni, T. D. Saini and A. A. Starobinsky, Mon. Not. Roy. Astron. Soc. \textbf{354}, 275 (2004).
\bibitem{Dabrowskisea} M.~P.~Dabrowski, K.~Marosek and A.~Balcerzak,
  Mem.\ Soc.\ Ast.\ It.\  {\bf 85}, 44 (2014)
  [arXiv:1308.5462 [astro-ph.CO]].

\bibitem{stoica} O. C. Stoica, Commun. Theor. Phys. \textbf{58}, 613 (2012);
G. Niz and N. Turok, Phys. Rev. D \textbf{75}, 026001 (2007); C.
Cattoen and M. Visser, Class. Quant. Grav. \textbf{22}, 4913 (2005);
J. M. M. Senovilla, Phys. Rev. Lett. \textbf{64}, 2219 (1990); G. L.
Murphy, Phys. Rev. D \textbf{8}, 4231 (1973).


\bibitem{Nojiri}S. Nojiri, S. D. Odintsov and S. Tsujikawa, Phys. Rev.D \textbf{71}, 063004 (2005).

\bibitem{type5} Y. Shtanov and V. Sahni, Class. Quant. Grav. \textbf{19}, L101 (2002);
M.P. D¸abrowski, T. Denkiewicz, Phys. Rev. D \textbf{79}, 063521
(2009); L. Fernandez-Jambrina, Phys. Rev. D \textbf{82}, 124004
(2010).

\bibitem{sudden} E.~I.~Guendelman and A.~B.~Kaganovich, Phys.\ Rev.\ D {\bf 75}, 083505 (2007)
  [gr-qc/0607111]; S. Nojiri, S.D. Odintsov, Phys. Lett. B \textbf{595},1 (2004); J.D. Barrow,
Class. Quant. Grav. \textbf{21}, 5619 (2004); L.P. Chimento, R.
Lazkoz, Mod. Phys. Lett. A \textbf{19}, 2479 (2004) ; J.D. Barrow,
A.B. Batista, J.C. Fabris, S. Houndjo, Phys. Rev. D \textbf{78},
123508 (2008); J.D. Barrow, S.Z.W. Lip, Phys. Rev. D \textbf{80},
043518 (2009); S. Nojiri, S.D. Odintsov, Phys. Rev. D \textbf{78},
046006 (2008); J.D. Barrow, S. Cotsakis, A. Tsokaros, Class. Quant.
Grav. \textbf{27}, 165017 (2010); J.D. Barrow, S. Cotsakis, A.
Tsokaros, [arXiv:1003.1027] (2010); P. Singh, Phys. Rev. D
\textbf{85}, 104011 (2012); T. Denkiewicz, M.P. D¸abrowski, H.
Ghodsi, M.A. Hendry, Phys. Rev. D \textbf{85}, 083527 (2012)

\bibitem{suddenb} M.P. Dabrowski, Phys. Lett. B \textbf{625}, 184 (2005).
\bibitem{Dabrowski:2004bz}
  M.~P.~Dabrowski,
  Phys.\ Rev.\ D {\bf 71}, 103505 (2005)
  [gr-qc/0410033].

\bibitem{type4b} J.D. Barrow, C.G. Tsagas, Class. Quant. Grav. \textbf{22}, 1563 (2005).


\bibitem{bbbb} V. Gorini, A.Y. Kamenshchik, U. Moschella, V. Pasquier,
Phys.\ Rev.\ D {\bf 69}, 123512 (2004); A.O. Barvinsky, C. Deffayet,
A.Yu. Kamenshchik, JCAP \textbf{05}, 034 (2010) [arXiv:0801.2063].


\bibitem{fjl} L.~Fernandez-Jambrina and R.~Lazkoz, Phys.\ Rev.\ D {\bf 70},
 121503 (2004)  [gr-qc/0410124].

\bibitem{type3} S. Nojiri and S. D. Odintsov, Phys. Rev. D \textbf{70}, 103522 (2004);
M. Bouhmadi-Lopez, P. F. Gonzalez-Diaz and P. Martin-Moruno, Phys.
Lett. B \textbf{659}, 1 (2008).


\bibitem{barrow} J. D. Barrow, Class. Quant. Grav. \textbf{21}, L79 (2004),Class. Quant. Grav. \textbf{21}, 5619 (2004);
S. Cotsakis and I. Klaoudatou, J. Geom. Phys. \textbf{55},
306(2005); K. Lake, Class. Quant. Grav. \textbf{21}, L129 (2004).



\bibitem{lr} P.H. Frampton, K.J. Ludwick, R.J. Scherrer, Phys. Rev. D \textbf{84}, 063003 (2011); P.H. Frampton,
K.J. Ludwick, S. Nojiri, S.D. Odintsov, R.J. Scherrer, Phys. Lett. B
\textbf{708}, 204 (2012)

\bibitem{sr}P.H. Frampton, K.J. Ludwick, R.J. Scherrer, Phys. Rev. D \textbf{85}, 083001 (2012)

\bibitem{srbr} M.~Bouhmadi-Lopez, A.~Errahmani, P.~Martin-Moruno, T.~Ouali and Y.~Tavakoli,
  Int.\ J.\ Mod.\ Phys.\ D {\bf 24}, 1550078 (2015)
  [arXiv:1407.2446 [gr-qc]].

\bibitem{grcr} L.~Fernandez-Jambrina, Phys.\ Rev.\ D {\bf 90}, 064014 (2014)  [arXiv:1408.6997 [gr-qc]].


\bibitem{ieos} I.~Brevik, V.~V.~Obukhov, K.~E.~Osetrin and A.~V.~Timoshkin, Mod.\ Phys.\ Lett.\ A {\bf 27}, 1250210 (2012)
 [arXiv:1210.4412 [gr-qc]]; I.~Brevik, V.~V.~Obukhov and A.~V.~Timoshkin, Astrophys.\ Space Sci.\  {\bf 344}, 275 (2013)
 [arXiv:1212.0391 [gr-qc]]; I.~Brevik, V.~Obukhov and A.~Timoshkin, TSPU Bulletin {\bf 128}, 42 (2012)  [arXiv:1303.5669
 [gr-qc]]; V.~V.~Obukhov, A.~V.~Timoshkin and E.~V.~Savushkin, Galaxies {\bf 1}, 107 (2013)  [arXiv:1309.4553 [gr-qc]].

\bibitem{cfr} I.~Brevik, A.~V.~Timoshkin and Y.~Rabochaya, Mod.\ Phys.\ Lett.\ A {\bf 28}, 1350172 (2013)  [arXiv:1311.5397 [gr-qc]].

\bibitem{cat} M.~Cataldo, N.~Cruz and S.~Lepe, Phys.\ Lett.\ B {\bf 619}, 5 (2005)  [hep-th/0506153].

\bibitem{bre} I.~Brevik, E.~Elizalde, S.~Nojiri and S.~D.~Odintsov, Phys.\ Rev.\ D {\bf 84}, 103508 (2011)  [arXiv:1107.4642 [hep-th]].

\bibitem{wcat} M.~Cataldo and P.~Meza, Phys.\ Rev.\ D {\bf 87}, no. 6, 064012 (2013)  [arXiv:1302.3748 [gr-qc]].

\bibitem{Campanelli} L. Campanelli, P. Cea, G. L. Fogli and A. Marrone, Phys. Rev. D {\bf 83}, 103503
(2011).
\bibitem{Rodriguez} R.~G.~Cai, Y.~Z.~Ma, B.~Tang and Z.~L.~Tuo,
  Phys.\ Rev.\ D {\bf 87}, no. 12, 123522 (2013); D. C. Rodrigues, Phys. Rev. D {\bf 77}, 023534 (2008);
J. Beltran Jimenez and A. L. Maroto, Phys. Rev. D {\bf 76}, 023003
(2007); L. Campanelli, P. Cea and L. Tedesco, Phys. Rev. Lett. {\bf
97}, 131302 (2006). 

\bibitem{Chimento} L.~P.~Chimento, Phys.\ Rev.\ D {\bf 68}, 023504 (2003),[gr-qc/0304033].

\bibitem{Belinski} Belinski V. A., Lifshits E. M., Khalatnikov I.
M., Usp. Fiz. Nauk, {\bf 102}, 463 (1970).

\bibitem{Hawking} S.W. Hawking and G.F.R. Ellis: The large scale structure of
space-time, Cambridge Monographs on Mathematical Physics (Cambridge
University Press, 1973).
\bibitem{Roy} S.R. Roy and S.K. Banerjee, Class. and Quantum Grav. {\bf 11}, 1943 (1995).

\bibitem{Bicak} J.~Bicak, Lect.\ Notes Phys.\  {\bf 540}, 1 (2000) [gr-qc/0004016]; K.C. Jacobs, Astrophys. J. {\bf 155}, 379 (1969).

\end{thebibliography}
\end{document}